\documentclass[twocolumn,superscriptaddress,amsmath,amssymb,showpacs,floatfix]{revtex4}
\usepackage{graphicx}
\usepackage{wasysym}
\usepackage{bm}

\begin{document}
\title{ Thermoelectric and thermal transport in bilayer graphene systems}
\author{R. Ma}
\affiliation{Faculty of Mathematics and Physics, Nanjing University
of Information Science and Technology, Nanjing 210044, China}
\affiliation{Department of Physics and Astronomy, California State
University, Northridge, California 91330, USA}
\author{L.  Zhu}
\affiliation{Theoretical Division and Center for Nonlinear Studies, Los Alamos National
Laboratory, Los Alamos, NM 87545, USA}
\author{L. Sheng}
\affiliation{National Laboratory of Solid State Microstructures and Department of Physics, Nanjing University, Nanjing 210093, China }
\author{M. Liu}
\affiliation{Department of Physics, Southeast University, Nanjing 210096, China}
\author{D.N Sheng}
\affiliation{Department of Physics and Astronomy, California State University, Northridge, California 91330, USA}

\begin{abstract}
We numerically study  the disorder effect on the thermoelectric and thermal transport for bilayer
graphene under a strong perpendicular magnetic field. In the unbiased case, we find that the thermoelectric transport has similar properties as in the monolayer graphene, i.e., the Nernst signal has a peak at the central Landau level (LL) with the value of the order of $k_B/e$ and changes sign near other LLs while the thermopower has an opposite behavior. We attribute this to the coexistence of particle and hole LLs around the Dirac point. When a finite interlayer bias is applied and a band gap is opened, it is found that the transport properties are consistent with those of a band insulator. We further study the thermal transport from electronic origins and verify the validity of the generalized Weidemann-Franz law.

\end{abstract}

\maketitle

\section{Introduction}
\label{sec:intro}

Thermoelectric transport properties of graphene have attracted much recent experimental~\cite{Kim09, Shi09, Ong08} and theoretical~\cite{CastroNeto09,DasSarma09,Fogelstrom07,Thalmeier07,Ting09,Zhu10} attention. The thermopower (the longitudinal thermoelectric response) and the Nernst signal (the transverse response) in the presence of a strong magnetic field are found to be large, reaching the order of the quantum limit $k_B/e$, where $k_B$ and $e$ are the Boltzmann constant and the electron charge, respectively~\cite{Kim09, Shi09, Ong08}. This has been attributed to the semi-metal type dispersion of graphene and/or in the vicinity of a quantum Hall liquid to insulator transition where the imbalance between the particle and hole types of carriers is significant. The thermoelectric effects are very sensitive to such an imbalance and become large in comparison with conventional metals.

In our previous study on graphene in the presence of disorder and an external magnetic field~\cite{Zhu10}, we have shown that its thermoelectric transport properties are determined by the interplay of the unique band structure, the disorder-induced scattering, the Landau quantization and the temperature. While the band structure and the magnetic field determine the Landau level (LL) spectrum, the broadening of each LL is controlled by the competition between disorder-induced scattering and the thermal activation. We find that all transport coefficients are universal functions of $W_L/E_F$ and $k_B T/E_F$ when both $W_L$ and $k_B T$ are much smaller than the Landau quantization energy $\hbar \omega_c$. Here $W_L$, $E_F$ and $T$  are the disorder-induced LL broadening, the Fermi energy and the temperature, respectively.  When $k_B T \ll W_L$, the thermoelectric conductivities vary as the density of states (and the particle-hole symmetry) is tuned by $E_F$ from the center of the LL to the mobility gap. When $k_B T \gg W_L$, thermal activation dominates and certain peak values for the thermopower $S_{xx}$ or the Nernst signal $S_{xy}$ reach universal numbers independent of the magnetic field or the temperature. While both $S_{xx}$ and $S_{xy}$ near high LLs ($\nu \neq 0$) have similar behaviors as those in two-dimensional (2D) semiconductor systems displaying the integer quantum Hall effect (IQHE)~\cite{Girvin82,Streda83,Jonson84,Oji84},  they rather have opposite behaviors around the central LL. $S_{xy}$ has a peak while $S_{xx}$ vanishes and changes sign at the Dirac point ($E_F=0$). We have further argued that the unique behavior at the central LL is due to the coexistence of particle and hole LLs. As protected by the particle-hole symmetry, the contributions from particle and hole LLs cancel with each other exactly in the thermopower but superpose in the Nernst signal. The results for such a tight-binding analysis are in good agreement with  experimental observations~\cite{Kim09, Shi09, Ong08}.

In this work, we extend our study to bilayer graphene which has two parallel graphene sheets stacked on top of each other as in 3D graphite (the AB or Bernal stacking). While some common features are observed related to LLs with the same underlying particle-hole symmetry, bilayer graphene also demonstrates some interesting and different properties from monolayer graphene~\cite{McCann06,Min07,Castro07,Oostinga08,Zhang09,Mak09}. The low energy dispersion of bilayer graphene can be effectively given by two hyperbolic bands $\epsilon_k \approx  \pm k^2/(2m^*)$ touching each other at the Dirac point ($E_F=0$), i.e., the electrons or holes have a finite mass $m^*$ which is in contrast to the massless excitations in monolayer graphene. Another important difference of bilayer graphene is the possibility to open up a band gap with a bias voltage, or a potential difference, applied between the two layers. This tunable gap system is advantageous to conventional semiconductor materials, making bilayer graphene more appealing from the point of view of applications. The thermoelectric transport properties of bilayer graphene are also expected to be interesting.  The the thermopower of bilayer graphene without a magnetic field has been considered~\cite{Hao10}.  It is shown that as the density of states is also of the pseudogap type without a biased voltage,  one expects that the relation for the thermopower  $S_{xx} \sim T/E_F$ continues to hold. In addition, it is found that the room-temperature thermopower with a bias voltage can be enhanced by a factor of 4 than monolayer graphene or unbiased bilayer graphene~\cite{Hao10}, making it a more promising candidate for future thermoelectric applications. Our study is to consider the thermopower and the Nernst effect under a magnetic field.

When an external magnetic field $B$ is applied, as in graphene and other IQHE systems, electron states of bilayer graphene are quantized into Landau levels. As the band dispersion changes, these LLs follow a different quantization sequence $E_n=\pm \sqrt{n(n-1)}\hbar\omega_c$ with $\omega_c \sim B$ rather than $\sqrt{B}$ for graphene. This has been confirmed by the theoretical~\cite{E.McCann06} and experimental~\cite{Novoselov06} studies on the quantum Hall effects, and further verified by our numerical calculation~\cite{Ma08}. Compared with graphene, though the massive nature of particles and hyperbolic dispersion are different, the existence of the central LL ($\nu=0$) and the associated chiral and particle-hole symmetries are preserved. Therefore, the study on the thermoelectric transport in bilayer graphene not only provides theoretical predictions for their properties, in particular, their dependence on disorder and magnetic field for this system, but also helps to verify our argument on the central LL that its unique behavior is due to the chiral and particle-hole symmetries associated with the Dirac point.

For such purposes, we carry out a numerical study of the thermoelectric transport in both unbiased and biased bilayer graphene. We focus on studying the effects of disorder and thermal activation on the broadening of LLs and the corresponding thermoelectric transport properties.  In the unbiased case, we indeed observe similar behaviors as in monolayer graphene  for the central LL. Both the longitudinal and the transverse thermoelectric conductivities are universal functions of $W_L/E_F$ and $k_BT/E_F$ and display different asymptotic behaviors in different temperature regions. The calculated Nernst signal has a peak at the central LL with heights of the order of $k_B/e$, and changes sign near other LLs, while the thermopower has an opposite behavior.  A higher peak value is obtained comparing to graphene due to the doubled degeneracy. This confirms our argument that as the particle and hole LLs coexist only in the central LL, the thermopower vanishes while the Nernst effect has a peak structure.  As before, we verify the validity of the semiclassical Mott relation, which is shown to hold in a wide range of temperatures. When a bias is applied between the two graphene layers, the thermoelectric coefficients exhibit unique characteristics quite different from those of unbiased case. Around the Dirac point, the transverse thermoelectric conductivity exhibits a pronounced valley with $\alpha_{xy}=0$ at low temperature, and the thermopower displays a very large peak. We show that these features are associated with a band insulator, due to the opening of a sizable gap between the valence and conductance bands in biased bilayer graphene. In addition, we have calculated the thermal transport properties of electrons for both unbiased and biased bilayer graphene systems. In the biased case, it is found that the transverse thermal conductivity displays a pronounced plateau with $\kappa_{xy}=0$, which is accompanied by a valley in $\kappa_{xx}$. This provides additional evidence for the band insulator behaviors.  We further compare the calculated thermal conductivities with those deduced from the Wiedemann-Franz law, to check the validity of this fundamental relation in graphene systems.

This paper is organized as follows. In Sec.\ II, we introduce the
model Hamiltonian. In Sec.\ III and Sec.\ IV, numerical results
based on exact diagonalization and thermoelectric transport
calculations are presented for unbiased and biased systems,
respectively. In Sec.\ V, numerical results for thermal transport
are presented. The final section contains a summary.

\section{Model and Methods}
\label{sec:model}

We consider a bilayer graphene sample consisting of two coupled
hexagonal lattices including inequivalent sublattices $A$, $B$ on
the bottom layer and $\widetilde{A}$, $\widetilde{B}$ on the top
layer. The two layers are arranged in the AB (Bernal)
stacking~\cite{S. B. Trickey,K. Yoshizawa}, where $B$ atoms are
located directly below $\widetilde{A}$ atoms, and $A$ atoms are the
centers of the hexagons in the other layer. Here, the in-plane
nearest-neighbor hopping integral between $A$ and $B$ atoms or
between $\widetilde{A}$ and $\widetilde{B}$ atoms is denoted by
$\gamma_{AB} =\gamma_{\widetilde{A}\widetilde{B}}=\gamma_{0}$. For
the interlayer coupling, we take into account the largest hopping
integral between $B$ atom and the nearest $\widetilde{A}$ atom
$\gamma_{\widetilde{A}B}=\gamma_{1}$, and the smaller hopping
integral between an $A$ atom and three nearest $\widetilde{B}$ atoms
$\gamma_{A\widetilde{B}}=\gamma_{3}$. The values of these hopping
integrals are taken to be $\gamma_{0}=3.16$ eV, $\gamma_{1}=0.39$
eV, and $\gamma_{3}=0.315$ eV, as same as in Ref.~\cite{Ma08}.

We assume that each monolayer graphene has totally $L_{y}$ zigzag
chains with $L_{x}$ atomic sites on each chain~\cite{Sheng06}. The
size of the sample will be denoted as $N=L_{x}\times L_{y}\times
L_{z}$, where $L_{z}=2$ is the number of monolayer graphene planes
along the $z$ direction. We model charged impurities in substrate,
randomly located in a plane at a distance $d$  from the
graphene sheet with  long-range Coulomb scattering
potentials~\cite{Adam08}. This type of disorder is known to give more
satisfactory results for  transport properties of graphene
in the absence of a magnetic field~\cite{Adam07}. When a magnetic
field is applied perpendicular to the bilayer graphene plane, the
Hamiltonian can be written in the tight-binding form

\begin{eqnarray}
H_0&=&-\gamma_{0}(\sum\limits_{\langle
ij\rangle\sigma}e^{ia_{ij}}c_{i\sigma}^{\dagger
}c_{j\sigma}+\sum\limits_{\langle
ij\rangle\sigma}e^{ia_{ij}}\widetilde{c}_{i\sigma}^{\dagger}\widetilde{c}_{j\sigma})\nonumber\\
&-&\gamma_{1}\sum\limits_{\langle
ij\rangle_1\sigma}e^{ia_{ij}}c_{j\sigma
B}^{\dagger}\widetilde{c}_{i\sigma\widetilde{A}}
-\gamma_{3}\sum\limits_{\langle
ij\rangle_3\sigma}e^{ia_{ij}}c_{i\sigma A}^{\dagger}\widetilde{c}_{j\sigma\widetilde{B}} +h.c.\nonumber\\
&+&\sum\limits_{i\sigma}w_{i}(c_{i\sigma}^{\dagger
}c_{i\sigma}+\widetilde{c}_{i\sigma}^{\dagger
}\widetilde{c}_{i\sigma}),
\end{eqnarray}

where $c_{i\sigma}^{\dagger}$($c_{i\sigma A}^{\dagger}$),
$c_{j\sigma}^{\dagger}$($c_{j\sigma B}^{\dagger}$) are creating
operators on $A$ and $B$ sublattices in the bottom layer, and
$\widetilde{c}_{i\sigma}^{\dagger}$($\widetilde{c}_{i\sigma\widetilde{A}}^{\dagger}$),
$\widetilde{c}_{j\sigma}^{\dagger}$($\widetilde{c}_{j\sigma\widetilde{B}}^{\dagger}$)
are creating operators on $\widetilde{A}$ and $\widetilde{B}$
sublattices in the top layer. $\sigma$ is a spin index. The sum
$\sum_{\langle ij\rangle\sigma}$ denotes the intralayer
nearest-neighbor hopping in both layers, $\sum_{\langle
ij\rangle_1\sigma}$ stands for interlayer hopping between the $B$
sublattice in the bottom layer and the $\widetilde{A}$ sublattice in
the top layer, and $\sum_{\langle ij\rangle_3\sigma}$ stands for the
interlayer hopping between the $A$ sublattice in the bottom layer
and the $\widetilde{B}$ sublattice in the top layer, as described
above. The magnetic flux per hexagon $\phi =\sum_{{\small
{\mbox{\hexagon}}}}a_{ij}=\frac{2\pi }{M}$ is proportional to the
strength of the applied magnetic field $B$, where $M$ is assumed to
be an integer and the lattice constant is taken to be unity. For
charged impurities,
$w_i=-\frac{Ze^2}{\epsilon}\sum_{\alpha}1/\sqrt{({\bf r}_i-{\bf
R}_{\alpha})^2+d^2}$, where $Ze$ is the charge carried by
an impurity, $\epsilon$ is the effective background lattice
dielectric constant, and ${\bf r}_i$ and ${\bf R}_{\alpha}$ are the
planar positions of site $i$ and impurity $\alpha$, respectively.
All the properties of the substrate (or vacuum in the case of
suspended graphene) can be absorbed into a dimensionless parameter
$r_s= Ze^2/(\epsilon \hbar v_F)$, where $v_F$ is the Fermi velocity
of the electrons. For simplicity, in the following calculation, we
fix the values of distance $d=1$nm and impurity density as $1\%$ of
the total sites, and tune $r_s$ to control the impurity scattering
strength.

For the biased system, the two graphene layers gain different
electrostatic potentials, and the corresponding energy difference is
given by $\Delta _g=\epsilon_2-\epsilon_1$ where
$\epsilon_1=-\frac{1}{2}\Delta_g$, and
$\epsilon_2=\frac{1}{2}\Delta_g$. The Hamiltonian can be written as:
$H=H_0+\sum\limits_{i\sigma}\epsilon_1
(c_{i\sigma}^{\dagger}c_{i\sigma}+\epsilon_2\widetilde{c}_{i\sigma}
^{\dagger}\widetilde{c}_{i\sigma}$). For illustrative purpose, a
relatively large asymmetric gap $\Delta_g=0.1\gamma_0$ is assumed,
which is experimentally achievable~\cite{Zhang09}.


In the linear response regime, the charge current in response to an
electric field or a temperature gradient can be written as  ${\bf J}
= {\hat \sigma} {\bf E} + {\hat \alpha} (-\nabla T)$, where ${\hat
\sigma}$ and ${\hat \alpha}$ are the electrical and thermoelectric
conductivity tensors, respectively.  These transport coefficients
can be calculated by Kubo formula once we obtain all the eigenstates
of the Hamiltonian (in our calculation, $\sigma _{xx}$ is obtained
based on the calculation of the Thouless number \cite{Ma08}). In
practice, we can first calculate the $T=0$ conductivities
$\sigma_{ji}(E_F)$, and then use the relation~\cite{Jonson84}
\begin{eqnarray}
\sigma_{ji}(E_F, T) &=& \int d\epsilon \,\sigma_{ji}(\epsilon)
\left ( - {\partial f(\epsilon) \over \partial \epsilon } \right), \nonumber \\
\alpha_{ji}(E_F, T) &=& {-1\over eT} \int d\epsilon\,
\sigma_{ji}(\epsilon) (\epsilon-E_F) \left ( - {\partial f(\epsilon)
\over \partial \epsilon } \right), \label{eq:conductance-finiteT}
\end{eqnarray}
to obtain the finite temperature electrical and thermoelectric
conductivity tensors. Here, $f(x) = 1/[e^{(x-E_F)/k_B T}+1]$ is the
Fermi distribution function. At low temperatures, the second
equation can be approximated as
\begin{equation}
\alpha_{ji}(E_F, T) =-\frac {\pi^2k_B^2T}{3e}\left. \frac
{d\sigma_{ji}(\epsilon, T)}{d\epsilon} \right|_{\epsilon =E_F},
\label{eq:Mott-relation}
\end{equation}
which is the semiclassical Mott relation~\cite{Jonson84,Oji84}. The
thermopower and Nernst signal can be calculated subsequently
from~\cite{footnote1}
\begin{eqnarray}
S_{xx} &=&  { E_x \over \nabla_x T} =  {\rho_{xx}\alpha_{xx}
- \rho_{yx}\alpha_{yx}}, \nonumber \\
S_{xy} &=& { E_y \over \nabla_x T} =  {\rho_{xx}\alpha_{yx}
+\rho_{yx}\alpha_{xx}}.
\label{eq:thermoelectric}
\end{eqnarray}


The thermal conductivity, measuring the magnitude of the thermal
currents in response to an applied temperature gradient, includes
electron and phonon contributions. In our numerical calculations,
phonon-derived thermal conductivity is omitted. The electronic
thermal conductivities $\kappa_{ji}$ at finite temperature assume the
forms~\cite{Oji84}
\begin{eqnarray}
\kappa_{ji}(E_F, T) &=& {1\over e^2T} \int d\epsilon\,
\sigma_{ji}(\epsilon) (\epsilon-E_F)^2 \left ( - {\partial
f(\epsilon) \over \partial \epsilon} \right) \nonumber\\
&-&T\alpha_{ji}(E_F, T)\sigma_{ji}^{-1}(E_F, T)\alpha_{ji}(E_F, T).
\label{eq:thermal conductivity}
\end{eqnarray}

For diffusive electronic transport in metals, it is well known
that the Wiedemann-Franz law is satisfied between the electrical
conductivity $\sigma$ and the thermal conductivity $\kappa$ of
electrons~\cite{Ziman}:
\begin{equation}
{\kappa\over \sigma T}=L , \label{eq:W-F law}
\end{equation}
where L is the Lorentz number and takes a constant
value:  $L={\pi^2\over 3}({k_B\over e})^2$.

\begin{figure}[tbp]
\par
\includegraphics[width=0.5\textwidth]{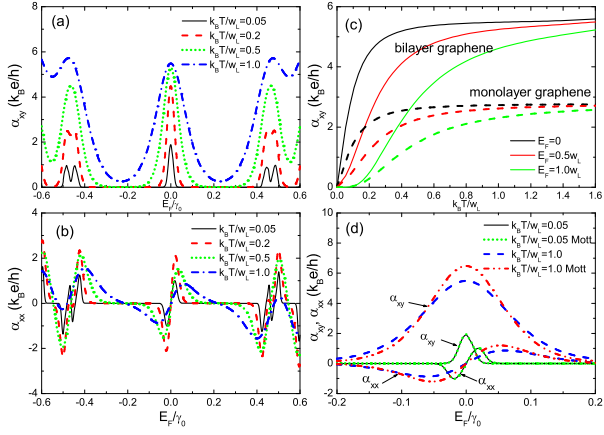}
\caption{(color online). Thermoelectric conductivities at finite
temperatures of bilayer graphene. (a)-(b) $\alpha_{xy}(E_F, T)$ and
$\alpha_{xx}(E_F,T)$ as functions of the Fermi energy at different
temperatures. (c) shows the temperature dependence of
$\alpha_{xy}(E_F,T)$ for monolayer and bilayer graphene. (d)
compares the results from numerical calculations and from the
generalized Mott relation at two characteristic temperatures,
$k_{B}T/W_L=0.05$ and $k_BT/W_L=1$. The system size is taken to be
$N=96\times48\times2$, magnetic flux $\phi=2\pi/48$, and disorder
strength $r_s =0.3$ (we consider uniformly distributed positive and
negative charged impurities within this strength) with $W_L/\gamma_0
= 0.0376$. } \label{fig.1}
\end{figure}

\section{Thermoelectric transport in unbiased bilayer graphene}
\label{sec:unbiased}

We first show  calculated thermoelectric
conductivities  at finite
temperatures for unbiased bilayer graphene. As seen from Fig.\ref{fig.1}(a)
and (b), the transverse thermoelectric conductivity $\alpha_{xy}$
displays a series of peaks, while the
longitudinal thermoelectric conductivity ${\alpha_{xx}}$ oscillates
and changes sign at the center of each LL. At low temperatures, the peak
of $\alpha_{xy}$ at the central LL is higher and narrower than others,
which indicates that the impurity scattering has less effect
on the central LL. These results
are qualitatively similar to those found in monolayer
graphene~\cite{Zhu10} due to the similar particle-hole symmetry in
both cases, but some obvious differences exist. Firstly, the
peak values of $\alpha_{xy}$ at the central LL is larger than that of
monolayer graphene. Secondly, at low temperature, $\alpha_{xy}$
splits around $E_F=\pm 0.46\gamma_0$, which can be understood as
due to the presence of $\nu=\pm 8$ Hall plateau by lifting subband degeneracy.
In Fig.\ref{fig.1}(c), we find that $\alpha_{xy}$ shows different
behavior depending on the relative strength of
temperature $k_{B}T$ and the width of the central LL $W_L$($W_L$ is
determined by the full-width at half-maximum of the $\sigma_{xx}$
peak). When $k_{B}T \ll W_L$ and $E_F \ll W_L$,
$\alpha_{xy}$ shows linear temperature dependence, indicating that
there is a small energy range where extended states dominate, and
transport fall into the semi-classical Drude-Zener regime.
When $E_F$ is shifted away from the Dirac point, the low temperature electron excitation is gapped
related to Anderson-localization.
When $k_{B}T$ becomes comparable to or greater than $W_L$, the
$\alpha_{xy}$ for all LLs saturates to a constant value $5.54 k_B
e/h$. This matches exactly the universal number $(\ln 2) k_B e/h$
predicted for the conventional IQHE systems in the case where
thermal activation dominates~\cite{Jonson84, Oji84}, with an
additional degeneracy factor $8$. The saturated value of
$\alpha_{xy}$ in bilayer graphene is exactly twice than that of
the monolayer graphene, as shown in Fig.\ref{fig.1}(c)
in accordance with the
eightfold degeneracy from valley, spin and layer degree of freedoms~\cite{Novoselov06,
E.McCann06}.

To examine the validity of the semiclassical Mott relation, we
compare the above results with those calculated from
Eq.(\ref{eq:Mott-relation}), as shown in Fig.\ref{fig.1}(d). The
Mott relation is a low-temperature approximation and predicts that
the thermoelectric conductivities have linear temperature dependence.
This is in agreement with our low-temperature results,
which proves that the semiclassical Mott
relation is asymptotically valid in Landau-quantized systems, as
suggested in Ref.~\cite{Jonson84}.

\begin{figure}[tbp]
\par
\includegraphics[width=0.5\textwidth]{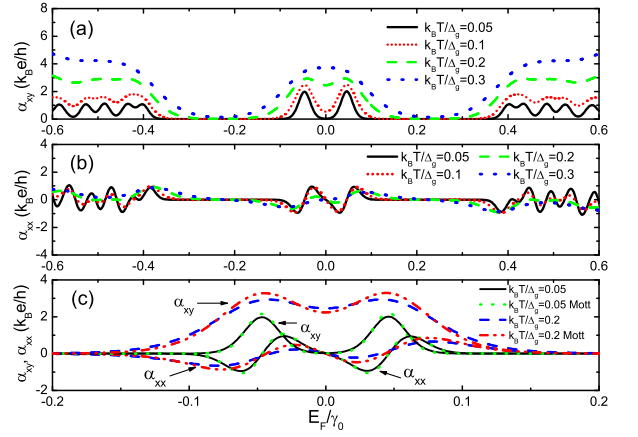}
\caption{(color online). Thermoelectric conductivities at finite
temperatures of biased bilayer graphene. (a)-(b) $\alpha_{xy}(E_F,
T)$ and $\alpha_{xx}(E_F,T)$ as functions of the Fermi energy at
different temperatures. (c) Compares the results from numerical
calculations and from the generalized Mott relation at two
characteristic temperatures, $k_{B}T/\Delta_g=0.05$ and
$k_BT/\Delta_g=0.2$. Here asymmetric gap $\Delta_g=0.1\gamma_0$. The
system size is taken to be $N=96\times48\times2$, magnetic flux
$\phi=2\pi/48$, and disorder strength $r_s =0.3$.} \label{fig.2}
\end{figure}

\begin{figure}[tbh]
\par
\includegraphics[width=0.5\textwidth]{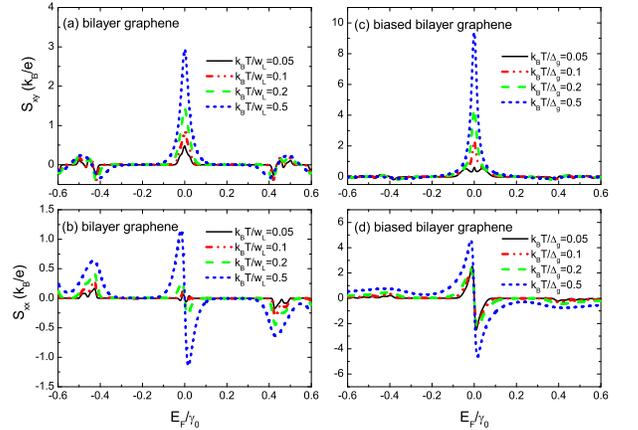}
\caption{ (color online). The thermopower $S_{xx}$ and the Nernst
signal $S_{xy}$  as functions of the Fermi energy in (a)-(b)bilayer
graphene, (c)-(d)biased bilayer graphene at different temperatures.
All parameters in this two systems are chosen to be the same as in
Fig. 1 and Fig. 2, respectively.} \label{fig.3}
\end{figure}

\section{Thermoelectric transport in biased bilayer graphene}
\label{sec:biased}

For biased bilayer graphene, we  show  results of
$\alpha_{xx}$ and $\alpha_{xy}$ at finite temperatures in Fig.
\ref{fig.2}.
Here we see that $\alpha_{xy}$ demonstrates a pronounced valley, in striking
contrast to the unbiased case with a peak at the particle-hole symmetric point $E_f=0$.
This behavior can  be understood as due to the split of the
valley degeneracy in the central LL by an opposite voltage bias added to the two
layers.  This is consistent with the
opening of a sizable gap between the valence and conduction bands.
More oscillations are observed
in the higher LLs comparing to the unbiased case, in consistent with
the further lifting of the LL degeneracy.
${\alpha_{xx}}$ oscillates and changes
sign around the center of each split LL.
In Fig.\ref{fig.2}(c), we also compare the above results with those
calculated from the semiclassical Mott relation using
Eq.(\ref{eq:Mott-relation}). Here the Mott relation is shown
to remain valid at low temperature.

We further calculate the thermopower $S_{xx}$ and the Nernst signal
$S_{xy}$ using Eq.\ (\ref{eq:thermoelectric}), which can be directly
determined in experiments by measuring the responsive electric
fields. In Fig. \ref{fig.3}(a)-(b), we show  results of $S_{xx}$
and $S_{xy}$ in unbiased bilayer graphene. As we can see, $S_{xy}$ ($S_{xx}$)
has a peak at the central LL (the other LLs), and changes sign near
the other LLs (the central LL), similar to the case of monolayer
graphene\cite{Zhu10}. This oscillatory feature has been observed
experimentally~\cite{Lee10}. In our calculation, the peak value of
$S_{xx}$ at $n=-1$ LL is found to be $14\mu V/K$ (note that $k_B/e = 86.17 \mu V/K$ )
for $k_BT=0.05W_L$
and $26\mu V/K$  for $k_BT=0.1W_L$, which is in good agreement with
the measured value~\cite{Lee10}. At zero energy, both $\rho_{xy}$
and $\alpha_{xx}$ vanish, leading to a vanishing $S_{xx}$. Around
the zero energy, because $\rho_{xx} \alpha_{xx}$ and
$\rho_{xy}\alpha_{xy}$ have opposite signs,
depending on their relative magnitudes, $S_{xx}$ could either increases or decreases when
 $E_F$ is
increased passing the Dirac point. In bilayer graphene, we find that
$S_{xx}$ is always dominated by $\rho_{xy}\alpha_{xy}$,
consequently, $S_{xx}$ decreases to negative value as $E_F$ passing
zero. We find that the peak value of $S_{xx}$ in the central LL is
$\pm 6\mu V/K$ at $k_BT=0.05W_L$. On the other hand, $S_{xy}$ has a
peak structure at zero energy, which is dominated by
$\rho_{xx}\alpha_{xy}$. The peak value is $42\mu V/K$ at
$k_BT=0.05W_L$. These results are in good agreement with the
experiments.


In Fig. \ref{fig.3}(c)-(d), we show the calculated $S_{xx}$ and
$S_{xy}$ in biased bilayer graphene system. As we can see, $S_{xy}$
($S_{xx}$) has a peak around zero energy (the other LLs), and
changes sign near the other LLs (zero energy). In our calculation,
$S_{xx}$ is dominated by $\rho_{xx}\alpha_{xx}$, which is different
from the unbiased bilayer graphene. At low temperature, the peak
value of $S_{xx}$ around zero energy keeps almost unchanged around
$\pm 181\mu V/K$, which is much larger than that of unbiased case.
With the increase of temperature, the peak height increases to $\pm
396\mu V/K$ at $k_BT=0.5\Delta_g$. Theoretical study~\cite{Hao10}
indicates that, the large magnitude of $S_{xx}$ is mainly a result
of the energy gap. On the other hand, $S_{xy}$ has a peak structure
around zero energy, which is dominated by $\alpha_{xy}\rho_{xx}$.
With $\sigma_{xx} \sim 2e^2/h$ near $E_F=0$, we find that the peak
height is $198\mu V/K$ at $k_BT=0.1\Delta_g$, which is larger than
that of unbiased case.

\begin{figure}[tbp]
\par
\includegraphics[width=0.5\textwidth]{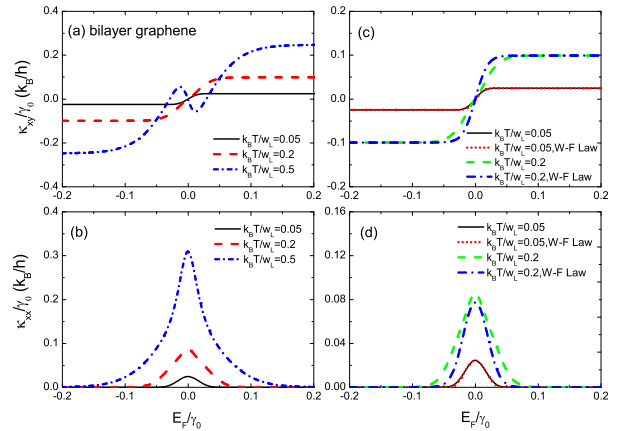}
\caption{(color online). (a)-(b) Thermal conductivities
$\kappa_{xy}(E_F,T)$ and $\kappa_{xx}(E_F,T)$ as functions of the
Fermi energy in bilayer graphene at different temperatures,
(c)-(d)Compares the thermal conductivity as functions of the Fermi
energy from numerical calculations and from the Wiedemann-Franz Law
at two characteristic temperatures. The parameters chosen here are
the same as in Fig. 1.} \label{fig.4}
\end{figure}

\begin{figure}[tbh]
\par
\includegraphics[width=0.5\textwidth]{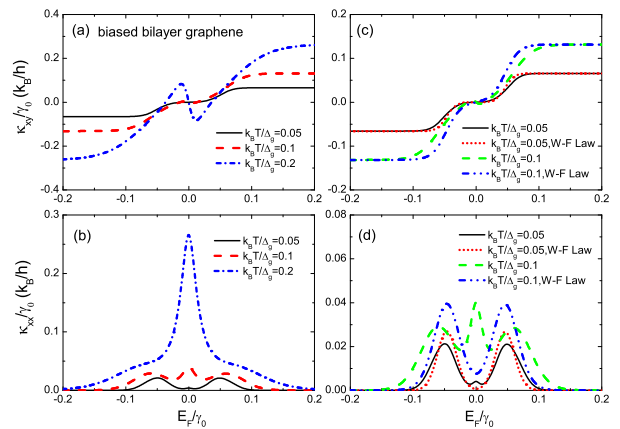}
\caption{ (color online). (a)-(b) Thermal conductivities
$\kappa_{xy}(E_F,T)$ and $\kappa_{xx}(E_F,T)$ as functions of the
Fermi energy in biased bilayer graphene at different temperatures,
(c)-(d)Compares the thermal conductivity as functions of the Fermi
energy from numerical calculations and from the Wiedemann-Franz Law
at two characteristic temperatures. The parameters chosen here are
the same as in Fig. 2. } \label{fig.5}
\end{figure}

\section{Thermal conductivity for unbiased and biased bilayer graphene systems}
\label{eq:thermal}

We now focus on  thermal conductivities. In Fig.\ {\ref{fig.4}, we
show  results of the transverse thermal conductivity $\kappa_{xy}$
and the longitudinal thermal conductivity $\kappa_{xx}$ for unbiased
bilayer graphene at different temperatures. As seen from
Fig.\ref{fig.4}(a) and (b), $\kappa_{xy}$ exhibits two flat plateaus
away from the center of the central LL. At low temperatures, the
transition between these two plateaus is smooth and monotonic, while
at higher  temperatures, $\kappa_{xy}$ exhibits an oscillatory
feature at $k_BT=0.5W_L$ between two plateaus. On the other hand,
$\kappa_{xx}$ displays a peak near the center of the central LL,
while its peak value increases quickly with $T$. To test the
validity of the Wiedemann-Franz Law, we compare the above results
with ones calculated from Eq.(\ref{eq:W-F law}) as shown in
Fig.\ref{fig.4}(c) and (d). The Wiedemann-Franz Law predicts that
the ratio of the thermal conductivity $\kappa$ to the electrical
conductivity $\sigma$ of a metal is proportional to the temperature.
This is in agreement with our low-temperature results, while
deviation is seen at higher $T$.

In Fig.\ {\ref{fig.5}, we show the calculated  thermal
conductivities $\kappa_{xx}$ and $\kappa_{xy}$ for biased bilayer
graphene. As seen from Fig.\ref{fig.5}(a) and (b), around the zero
energy, a flat region with $\kappa_{xy}=0$ is found at low
temperatures, which is accompanied by a valley in $\kappa_{xx}$.
These features are clearly in contrast to those of unbiased case due
to the asymmetric gap between the valence and conduction bands. When
temperature increases to $k_BT=0.2\Delta_g$, the plateau with
$\kappa_{xy}=0$ disappears, while $\kappa_{xx}$ displays a large
peak. In Fig.\ref{fig.5}(c) and (d), we also compare  above results
with those calculated from the Wiedemann-Franz Law using
Eq.(\ref{eq:W-F law}).
Due to the presence of energy gap, we find that
the Wiedemann-Franz Law is not
valid in the biased bilayer graphene.

\section{Summary}
\label{sec:sum}

In summary, we have numerically investigated the thermoelectric and
thermal transport in unbiased bilayer graphene based on the
tight-binding model in the presence of both disorder and magnetic
field. We find that the thermoelectric conductivities display
different asymptotic behaviors depending on the ratio between the
temperature and the width of the disorder-broadened Landau levels
(LLs), similar to those found in monolayer graphene. In the high
temperature regime, the transverse thermoelectric conductivity
$\alpha_{xy}$ saturates to a universal quantum value $5.54 k_B e/h$
at the center of each LL, and it has a linear temperature dependence
at low temperatures. The calculated Nernst signal $S_{xy}$ has a
peak at the central LL with heights of the order of $k_B/e$, and
changes sign at the other LLs, while the thermopower $S_{xx}$ has an
opposite behavior. These results are in good agreement with the
experimental observation\cite{Lee10}. The validity of the
semiclassical Mott relation between the thermoelectric and
electrical transport coefficients is verified in a range of
temperatures. The calculated transverse thermal conductivity
$\kappa_{xy}$ exhibits two plateaus away from the band center. The
transition between this two plateaus is continuous, which is
accompanied by a pronounced peak in longitudinal thermal
conductivity $\kappa_{xx}$. The validity of the Wiedemann-Franz Law
between the thermal conductivity $\kappa$ and the electrical
conductivity $\sigma$ is only verified at very low temperatures.

We further discuss the thermoelectric transport of biased bilayer
graphene. When a bias is applied to the two graphene layers, the
thermoelectric coefficients exhibit unique characteristics different
from those of unbiased case. Around the Dirac point, transverse
thermoelectric conductivity exhibits a pronounced valley with
$\alpha_{xy}=0$ at low temperatures, and the thermopower displays a
large magnitude peak. Furthermore, the transverse thermal
conductivity has a pronounced plateau with $\kappa_{xy}=0$, which is
accompanied by a valley in $\kappa_{xx}$. These are in consistent
with
the opening of sizable gap between the valence
and conductance bands in biased bilayer graphene.

We mention that in our numerical calculations, the magnetic field is
much stronger than the ones one can realize in the experimental
situation, as limited by current computational capability. However, the
asymptotic behaviors we obtained is robust and applicable to weak
field limit since it is determined by the topological property of
the energy band as clearly established for monolayer
graphene~\cite{Zhu10}.

\acknowledgments
This work is supported by the
DOE Office of Basic Energy Sciences under grant DE-FG02-06ER46305 (RM, DNS),
 and the U.S. DOE through the LDRD program at LANL (LZ), the NSF Grant DMR-0906816 (RM).
We also thank partial support from Princeton MRSEC Grant DMR-0819860, the NSF instrument grant
DMR-0958596 (DNS),  the NSFC Grant No. 10874066, the National Basic Research Program of China
under Grant Nos. 2007CB925104 and 2009CB929504 (LS), and the doctoral
foundation of Chinese Universities under Grant No.
20060286044 (ML). \\


\begin{references}
\bibitem{footnote1} Different literatures may have a sign difference due to different conventions.
\end{references}
\end{document}